\documentclass[12pt,pra,aps,amssymb,amsfonts,amsmath,tightenlines]{revtex4}

\usepackage[dvips]{graphicx}
\usepackage{dsfont, color} 

\newcommand{\ri}{\mathrm{i}}
\newcommand{\re}{\mbox{$\rm e$}}
\newcommand{\rd}{\mbox{$\rm d$}}
\newcommand{\half}{\mbox{$\textstyle \frac{1}{2}$}}
\def\bitcoinB{\leavevmode
  {\setbox0=\hbox{\textsf{\scriptsize{B}}}%
    \dimen0\ht0 \advance\dimen0 0.2ex
    \ooalign{\hfil \box0\hfil\cr
      \hfil\vrule height \dimen0 depth.2ex\hfil\cr
    }%
  }%
}

\begin{document}

\title{Theory of Cryptocurrency Interest Rates}
\author{Dorje C. Brody$^1$, Lane P. Hughston$^{2}$ and Bernhard K. Meister$^{3}$}
\affiliation{$^1$Department of Mathematics, University of Surrey, Guildford GU2 7XH, UK\\ 
$^{2}$Department of Computing, Goldsmiths College, University of London, New Cross, London SE14 6NW, UK
\\ $^{3}$School of Management, Swansea University, Swansea SA1 8EN, UK}

\date{\today}

\begin{abstract}
\noindent 
A term structure model in which the short rate is zero is developed as a candidate for a theory of cryptocurrency interest rates. The price processes of crypto discount bonds are worked out, along with expressions for the instantaneous forward rates and the prices of interest-rate derivatives. The model admits functional degrees of freedom that can be calibrated to the initial yield curve and other market data. Our analysis suggests that strict local martingales can be used for modelling the pricing kernels associated with virtual currencies based on distributed ledger technologies.

\begin{center}
{\scriptsize {\bf Keywords: cryptocurrencies, distributed ledger technologies, blockchains, \\interest rate models, pricing kernels, foreign exchange, derivatives.
} }
\end{center}
\end{abstract}

\maketitle


\noindent \textbf{1. Introduction} \\

\noindent Over a  decade has passed since the release of the white paper 
establishing bitcoin --- the progenitor of the burgeoning open-source 
cryptocurrency movement \cite{SN}. In the intervening years the ability of 
digital currencies to flourish with market capitalizations in the billions of 
dollars without the backing of sovereign states has been widely publicized. 
One of the key factors that has to this date hindered cryptocurrencies from 
reaching the mainstream and expanding beyond fringe applications to link
with the real economy in significant ways, despite numerous preliminary attempts, 
has been the inability of regulated trading consortiums to 
develop broad platforms for cryptocurrency derivatives and 
structured products, and more generally to put in place the robust infrastructure 
needed to support financial markets analogous to those we take for granted in 
connection with sovereign currencies. In economies based on well-established 
sovereign currencies, interest rate derivatives are central to the functioning 
of financial markets. A statistical bulletin published in December 2018 by the 
Basel-based Bank of International Settlement  illustrates how the notional 
amount for fixed-income derivatives trumps every other category by a multiple 
of five or more \cite{BIS}. For an interest-rate derivatives market to function, 
products need to be priced, payoffs need to be replicated, and positions need 
to be hedged --- and for pricing, replicating, and hedging, it is essential that 
financial institutions, regulators, and other market participants  should have at 
their disposal a diverse assortment of serviceable interest-rate models that are well 
adapted to characteristics of the currencies in which the instruments being 
traded are based. When it comes to cryptocurrencies, this requirement 
immediately poses a challenge, since cryptocurrencies by their nature offer 
no short-term interest, and it seems  impossible {\it prima facie} that one should 
be able to build an interest rate model for which the short rate is identically zero. 
In the case of bitcoin, the regularly-updated blockchain represents the 
distributed ledger and therefore holdings in the cryptocurrency. The 
technical details of the updating process are highly involved, but do not 
impinge upon the general statement that holdings of bitcoin as 
recorded in the ledger do not earn interest. No central bank exists. New 
coins are not issued to existing holders of the coins, but instead are awarded 
to ``miners'' successfully solving numerical puzzles necessary to update the 
blockchain. Other prominent cryptocurrencies provide variations on this 
general theme of not paying by 
accommodating additional functionality. For example, ethereum, the second 
largest cryptocurrency,  incorporates distributed computing.

What are the implications for interest rate modelling? To begin, it may be 
helpful if we recall what is generally meant by a ``conventional" interest 
rate model. This will allow us to identify some of the differences between 
the conventional theory and the interest rate theory required for cryptocurrencies. 
Such conventional models exist in abundance, and include, for instance, most 
of the examples mentioned in \cite{LPH1, LPH2}. One typically assumes the 
existence of a probability space $(\Omega, \mathcal F, \mathbb P)$ with an 
associated filtration 
$\{\mathcal F_t\}_{t\geq0}$ satisfying the usual conditions, upon which an adapted short-rate process 
$\{r_t\}_{t\geq0}$ is defined. Here, time $0$ denotes the present. By a unit discount bond we mean a financial instrument that delivers a single cash flow of one unit of currency at time $T$ and derives its value entirely from that cash flow.  The price $P_{tT}$ of 
a $T$-maturity unit discount bond at time $t < T$  is then
given by 
a conditional expectation of the form
\begin{eqnarray}
P_{tT}=  {\mathbb E}^{\mathbb Q}\left[ \left. \
\exp\left( -\int_t^T r_s \, \rd s \right) \right| \mathcal F_t\right], 
\label{traditional bond price} 
\end{eqnarray} 
taken under a suitably-specified risk-neutral measure $\mathbb Q$ that is equivalent (in the sense of agreeing on null sets) to the physical measure $\mathbb P$. Glancing at equation (\ref{traditional bond price}) one might conclude that it is not possible to have a nontrivial discount-bond system 
if $r_t=0$ for all $t \geq 0$, and indeed this is true in ``conventional'' models. For example, in the HJM model \cite{HJM}, the instantaneous forward rates, which determine the discount bond prices via the relation 
\begin{eqnarray}
P_{tT} =  \exp\left(-\int_t^T f_{tu} \, \rd u\right),
\end{eqnarray} 
for $t < T$, are given by 
\begin{eqnarray}
f_{tT}=  \frac{1}{P_{tT}} \, {\mathbb E}^{\mathbb Q}\left[ \left. r_T \,  \exp\left( -\int_t^T r_s \, \rd s \right) \right| \mathcal F_t\right]. 
\label{instantaneous forward rates} 
\end{eqnarray} 
Thus $f_{tT}$ is the forward price per unit notional, made at time $t$, for purchase of the rights to a cash flow of the amount $r_T$ per unit notional at time $T$. It follows that if the short rate vanishes then so do the instantaneous forward rates, in which case the discount bond system trivializes and takes the form $P_{tT} = 1$ for all $t<T$.  The same conclusion follows directly from \eqref{traditional bond price}. Thus if the short rate vanishes in a conventional interest rate model it follows immediately that so do all the term rates. 

Nevertheless, just because the short rate vanishes we are not necessarily forced to conclude that the discount bond system is trivial. Beginning with the work of Constantinides \cite{C}, finance theorists have learned to think about interest-rate modelling in a more general way, in terms of so-called pricing kernels. The pricing kernel method avoids some of the technical issues that arise with the introduction of the risk neutral measure and the selection of a preferred numeraire asset in the form of a money market account, and at the same time it leads to interesting new classes of interest rate models (see, e.g.,~\cite{BH1, HR, BH2, HK, Rogers} and references cited therein). 
By a pricing kernel, we mean an $\{{\mathcal F_t}\}$-adapted c\`adl\`ag  semimartingale $\{\pi_t\}_{t\geq0}$  satisfying {\rm(a)} $\pi_t >0$ for $t\geq0$, {\rm(b)} ${\mathbb E}\, [\,\pi_t\,] < \infty$ for
$t\geq0$, and {\rm(c)} $\liminf_{t\to\infty} 
{\mathbb E}\,[\pi_t]=0$,  with the property that  if an asset with value process $\{S_t\}_{t\geq0}$ delivers a bounded $\mathcal F_T$-measurable cash flow $H_T$ at time $T$ and derives its value entirely from that cash flow, 
then   
\begin{eqnarray}
S_{t} =  {\mathds 1}_{\{ t< T\} } \, \frac{1}{\pi_t}\, {\mathbb E} \left[ \left. \pi_T H_T \right| \mathcal F_t \right] . 
\label{pricing formula}
\end{eqnarray}
Here we write $\mathbb E$ for expectation under $\mathbb P$. We adopt the convention that the value of such an asset drops to zero at the instant the cash flow occurs. Thus $\lim_{t \to T} S_t = H_T$, whereas $S_t = 0$ for $t \geq T$.  This is in keeping with the usual analysis of stock prices when a stock goes ex-dividend, and respects the requirement that the price process should be c\`adl\`ag.  Our assumptions imply that $\pi_T H_T$ is integrable and that
$\{ \pi_t S_{t} + {\mathds 1}_{\{ t \geq T \} } \, \pi_T H_T \}_{t\geq 0}$ is a uniformly integrable martingale. 
The existence of an established pricing kernel is equivalent, in a broad sense, to what we mean by market equilibrium and the absence of arbitrage. In fact, it can be shown under rather general conditions \cite{JR} that with a few reasonable assumptions any pricing formula for contingent claims takes the form  (\ref{pricing formula}). Then if $H_T=1$, we obtain an expression for the price at time $t$ of a bond 
that pays one unit of currency at $T$, given by
\begin{eqnarray} 
P_{tT} =  {\mathds 1}_{\{ t< T\} } \, \frac{1}{\pi_t}\, {\mathbb E} \left[ \left. \pi_T \right| \mathcal F_t \right] . 
\label{bond price} 
\end{eqnarray}
To be sure, models of the type  represented by formula (\ref{traditional bond price}) can be obtained as instances of models of the type  represented by formula (\ref{bond price}), but it is not the case that all interest rate models are of type (\ref{traditional bond price}). As we shall demonstrate, models of type (\ref{bond price}) can be constructed for which the unit of conventional currency is replaced by a unit of cryptocurrency, and in such a way that we arrive at a 
nontrivial interest rate model for which the cryptocurrency condition $r_t=0$ is satisfied for all $t\geq 0$ and yet for which term rates are non-vanishing. 

The present paper explores a class of such models achieved by 
allowing the crypto pricing kernel to be a strict local martingale. The reasoning behind this proposal is as follows. The pricing kernel methodology requires that the price $\{S_t\}$ of a non-dividend-paying asset should have the property that the product $\{\pi_t S_t\}$ should be a martingale. Now, suppose the market admits a unit-initialized absolutely continuous money market account with value process  $\{B_t\}$ of the form
\begin{eqnarray}
B_{t}=\exp \left( \int_{0}^{t}r_{s} \, \rd s\right).
\label{money market account}
\end{eqnarray}
Then the process $\{\Lambda_t\}$ defined by $\Lambda_t = \pi_t B_t$ is a martingale, and the pricing kernel as a consequence is given by
\begin{eqnarray}
\pi_{t}=\exp{\left( -\int_{0}^{t}r_{s} \, \rd s\right) } \, \Lambda_t \,.
\end{eqnarray}
In that case, if we introduce the Radon-Nikodym derivative defined by  
\begin{eqnarray}
\left . \frac{\rd \mathbb Q} { \rd \mathbb P}\right |_{\mathcal F_t}  = \Lambda_t \, ,
\end{eqnarray}
we can make a change the measure in (\ref{pricing formula}), and we are led to the well-known risk-neutral valuation formula 
\begin{eqnarray}
S_{t} =  {\mathds 1}_{\{ t< T\} } \, {\mathbb E}^{\mathbb Q} \left[ \left. \exp\left( -\int_t^T r_s \, \rd s \right) H_T \right| \mathcal F_t\right]. 
\label{risk neutral valuation formula}
\end{eqnarray}
Then if we set $H_T = 1$ we recover the class of ``conventional" interest rate models defined by a bond price of the form (\ref{traditional bond price}) along with the money market account (\ref{money market account}).  
 
In the class of models we consider for crypto interest rates, which will be introduced in Section 2,  we exclude the existence of an instantaneous money-market asset from the model altogether. This can be achieved by choosing the pricing kernel to be a strict local martingale. This implies that the short rate vanishes and hence that the process $\{B_t\}$ defined by \eqref{money market account} is constant. But if $\{B_t\}$ is constant, then $\{\pi_t B_t\}$ is not a martingale, and thus there is no money-market asset. We consider a model in which the pricing kernel is given by the reciprocal of a Bessel process of order three. This reciprocal process, 
introduced in \cite{JH}, is a well-known example of a strict local martingale 
\cite{RY,MY,PP,FP}, and has the advantage of being 
highly tractable. 
The idea that this process can be used as a pricing kernel 
appears in \cite{HK}, where it is recognized that the resulting interest rate model does not admit a representation for the bond price
in the form (\ref{traditional bond price}). 
Here we develop a model of this type in detail in the context of cryptocurrency bonds. 
In Section 3 we derive explicit expressions for the discount bond system and 
the various associated rates. The results are applied in Section 4 to obtain pricing 
formulae for digital options on discount bonds and caplets on simple crypto 
rates. 
In Section 5, we introduce a class of related models based on Bessel($n$) processes with $n \geq 4$, and the fourth order model is
worked out in detail. In Section 6, we introduce a class of models based on a complexification of the Bessel(3) process.  We conclude in Section 7 with some remarks about options on crypto exchange rates.\\

\noindent \textbf{2. A model of no interest} \\

\noindent 
The pricing kernel formalism allows us to identify where the theory of cryptocurrencies deviates from the conventional one: namely, there 
is no money market account. But how 
is it possible to construct an interest rate model without a money market account? This 
apparent impossibility in the context of a conventional interest 
rate theory is nonetheless possible in a pricing kernel framework. 
The argument is as follows. First, we observe, by virtue of (\ref{bond price}), that a necessary and sufficient condition for the bond price to be a decreasing function of $T$ for any fixed $t$ such that $t<T$ is that $\{\pi_t\}$ should be a supermartingale. This implies that the interest rate system associated with $\{\pi_t\}$ is non-negative. Now, cryptocurrencies are by their nature storable assets, with negligible storage costs, and hence by a standard arbitrage argument cannot be borrowed at a negative rate of interest. Thus it is reasonable to assume that the crypto pricing kernel is a supermartingale. In any case, in what follows we shall make that assumption. We recall that a semimartingale is a local martingale if for any 
increasing sequence of stopping times $\{\tau_n\}_{n \in {\mathds N}}$ 
with $\lim_{n \rightarrow \infty} \tau_n = \infty$, the stopped process is a martingale for each value of $n$. A 
strict local martingale then is a local martingale that is not a true martingale. Now, it is well known that a positive local martingale is necessarily a supermartingale.  Thus, if we allow for 
the possibility that the 
pricing kernel is a strict local martingale, then the 
positivity of the pricing kernel implies that it is a supermartingale with vanishing drift. 
This suggests  that the cryptocurrency interest-rate term structure can be modelled by letting 
the pricing kernel be a strict local martingale. 

As an illustration,  we examine an interest rate model based on a pricing kernel given by
the reciprocal of the Bessel process of order three. Specifically, the pricing kernel is constructed as follows. Let $\{W_t^{(1)},W_t^{(2)},
W_t^{(3)} \}_{t\ge0}$ be three independent standard Brownian motions on a probability space 
$(\Omega, \mathcal F, \mathbb P)$ with filtration 
$\{\mathcal F_t\}_{t\geq0}$, and define 
\begin{eqnarray}
X_t = \int_0^t \sigma_s \rd W_s^{(1)}, \quad 
Y_t = \int_0^t \sigma_s \rd W_s^{(2)}, \quad 
Z_t = \int_0^t \sigma_s \rd W_s^{(3)} , 
\label{Brownian motions}
\end{eqnarray}
where $\{\sigma_t\}_{t \geq 0}$ is a deterministic function which we take to be bounded, strictly positive, and left-continuous. We then define a model 
for the pricing kernel by setting  
\begin{eqnarray} 
\pi_t = \frac{1}{\sqrt{(X_t-a)^2+(Y_t-b)^2+(Z_t-c)^2}} \, ,  
\label{pi}
\end{eqnarray}
where $a,b,c$ are constants, not all equal to zero. The initial condition 
$\pi_0=1$ requires that $a^2+b^2+c^2=1$, and rotational symmetry implies 
that the vector $(a,b,c)$ can lie on any point on the unit sphere in $ {\mathds R}^3$. 
We note that the function $u: {\mathds R}^3 \rightarrow {\mathds R}^+\cup{\infty}$ 
defined by
\begin{eqnarray}
u(x,y,z) = \frac{1}{\sqrt{(x-a)^2+(y-b)^2+(z-c)^2}} 
\end{eqnarray}
is a solution of Laplace's equation
\begin{eqnarray}
\frac{\partial^2}{\partial x ^2} u(x,y,z)+ \frac{\partial^2}{\partial y ^2} u(x,y,z)+ \frac{\partial^2}{\partial z ^2}u(x,y,z)
= 0 
\end{eqnarray}
on ${\mathds R}^3 \backslash (a,b,c)$. One can think of $\{u(x,y,z)\}$ as the Coulomb potential generated by a unit charge situated at the point $(a,b,c)$. Then we have $\pi_t = u(X_t, Y_t, Z_t)$, and using the fact
that the three Brownian motions are independent, we find that  the drift of $\{\pi_t\}$ vanishes and that  $\{\pi_t\}$ satisfies the dynamical equation 
\begin{eqnarray} 
\rd \pi_t = - \sigma_t \, \pi_t^2 \, \rd W_t \, , 
\label{Bessel-3 dynamics} 
\end{eqnarray}
where the process $\{W_t\}_{t\geq 0}$ defined by the 
relation 
\begin{eqnarray}
\rd W_t = \pi_t \left[ (X_t-a)\rd W_t^{(1)} + (Y_t-b)\rd W_t^{(2)} + 
(Z_t-c)\rd W_t^{(3)} \right] 
\end{eqnarray} 
is a standard Brownian motion. The interpretation of  (\ref{Bessel-3 dynamics}) is  as follows. Suppose we consider a generic market model in which the pricing kernel is a strictly positive Ito process driven by a Brownian motion  $\{W_t\}$ and satisfies a dynamical equation of the form 
\begin{eqnarray} 
\rd \pi_t =  \alpha_t \, \pi_t \, \rd t + \beta_t \, \pi_t \,  \rd W_t    \, .
\end{eqnarray}
Let  $\{S_t\}$ be the price of a non-dividend-paying asset driven by the same Brownian motion. Then  $\pi_t S_t = M_t$ for some positive martingale $\{M_t\}$ driven by $\{W_t\}$. Let  the dynamics of $\{M_t\}$ be given by $\rd M_{t}=\nu_{t}M_t \, \rd W_{t}$. A calculation using Ito's formula then shows that
\begin{eqnarray} 
\rd S_t =   [- \alpha_t -\beta_t  (\nu_t - \beta_t )]\, S_t \, \rd t  + (\nu_t - \beta_t )\, S_t \, \rd W_t \, . 
\end{eqnarray}
Thus if one writes $r_t = -\alpha_t$,  $\lambda_t = -\beta_t$ and $\sigma_t = \nu_t - \beta_t$, it follows that  the dynamics of the asset price takes the familiar form 
\begin{eqnarray} 
\rd S_t =   (r_t + \lambda_t  \sigma_t )\, S_t \, \rd t  + \sigma_t \, S_t \, \rd W_t \, . 
\label{asset dynamics}
\end{eqnarray}
We see that $r_t$ is the short rate of interest, that $\lambda_t$ is the market price of risk,
that $\sigma_t$ is the volatility of the asset, and that the pricing kernel satisfies 
\begin{eqnarray} 
\rd \pi_t = -r_t \pi_t \, \rd t - \lambda_t \pi_t \,  \rd W_t \, .
\label{pricing kernel dynamics}
\end{eqnarray}
Combining (\ref{Bessel-3 dynamics}) and (\ref{pricing kernel dynamics}), we deduce that in the Bessel(3) model for the pricing kernel the short rate satisfies
$r_t=0$ for all $t\geq0$, and the market price of risk is given by $\lambda_t=\sigma_t \pi_t$. The  vanishing of the short rate does not, however, imply the vanishing of rates of finite tenor, such as Libor rates and swap rates, as we shall see below. \\ \\

\noindent \textbf{3. Discount bonds and yields}\\

\noindent To work out the bond price process we shall be using the pricing 
formula (\ref{bond price}): 
\begin{eqnarray}
P_{tT} = \frac{1}{\pi_t} {\mathbb E}_t\left[ 
\frac{1}{\sqrt{(X_T-a)^2+(Y_T-b)^2+(Z_T-c)^2}} \right] ,   
\end{eqnarray}
where $\mathbb E_t[\, \cdot \,] := \mathbb E[\left.\, \cdot \,\right | \mathcal F_t]$. Writing 
$X_T-a=(X_T-X_t)+(X_t-a)$, and similarly for $Y_T-b$ and $Z_T-c$, we 
observe that the increments $X_T-X_t$, $Y_T-Y_t$, and $Z_T-Z_t$ are 
independent of ${\mathcal F}_t$, whereas $X_t-a$, $Y_t-b$, and $Z_t-c$ are 
${\mathcal F}_t$-measurable. Thus, defining $X=X_T-a$, $Y=Y_T-b$, and 
$Z=Z_T-c$, then conditionally on ${\mathcal F}_t$ we have $X\sim N
(X_t-a,\Sigma_{tT})$, $Y\sim N(Y_t-b,\Sigma_{tT})$ and 
$Z\sim N (Z_t-c,\Sigma_{tT})$, where 
\begin{eqnarray} 
\Sigma_{tT} = \int_t^T \sigma^2_s \, \rd s 
\end{eqnarray} 
is the conditional variance of the random variables $X$, $Y$, and $Z$. 
If we define the vectors ${\boldsymbol R}=(x,y,z)$ and ${\boldsymbol\xi_t}=
(X_t-a,Y_t-b,Z_t-c)$, and their squared norms $R^2={\boldsymbol R}\cdot
{\boldsymbol R}$ and $\xi_t^2={\boldsymbol\xi_t}\cdot{\boldsymbol\xi_t}$, then 
the bond price is given by 
\begin{eqnarray}
P_{tT} = \frac{1}{\pi_t} 
\frac{1}{(\sqrt{2\pi \Sigma_{tT}}\,)^3} \int_{{\mathds R}^3} 
\frac{1}{R}\, \re^{-\frac{1}{2} \Sigma_{tT}^{-1} \, |{\boldsymbol R}-{\boldsymbol\xi_t}|^2} 
\rd^3 {\boldsymbol R} \,  . 
\label{eq:15.0} 
\end{eqnarray} 
Thus using spherical representation 
\begin{eqnarray}
\rd^3 {\boldsymbol R}=R^2 \sin\theta\, \rd R \,\rd\theta\, \rd\phi
\end{eqnarray}
for the volume element in ${\mathds R}^3$ we deduce that  
\begin{eqnarray}
P_{tT} &=& \frac{1}{\pi_t} 
\frac{1}{\sqrt{2\pi}\Sigma_{tT}^{3/2}} 
\int_0^\infty  R^2\, \frac{1}{R} 
\int_0^\pi \sin\theta\, \re^{ -\frac{1}{2} \Sigma_{tT}^{-1} \,
(R^2-2R\xi_t\cos\theta+\xi_t^2)} \rd\theta \, \rd R
\nonumber \\ 
&=& \frac{1}{\pi_t} 
\frac{1}{\sqrt{2\pi}\Sigma_{tT}^{3/2}}  \, \re^{-\frac{1}{2} \Sigma_{tT}^{-1} \,\xi_t^2}
\int_0^\infty R\, \re^{  -\frac{1}{2} \Sigma_{tT}^{-1} \,R^2} \int_0^\pi 
\sin\theta \,  \re^{ R\xi_t\cos\theta/\Sigma_{tT}} \rd\theta \, \rd R \nonumber \\ 
&=& \frac{1}{\pi_t} \frac{1}{\sqrt{2\pi} \Sigma_{tT}^{3/2}}\, 
\re^{-\frac{1}{2} \Sigma_{tT}^{-1} \, \xi_t^2}
\int_0^\infty R\, \re^{  -\frac{1}{2} \Sigma_{tT}^{-1} \,R^2} \left[ \left. 
\frac{-\re^{ R\xi_t\cos\theta/\Sigma_{tT}}}{R\xi_t/\Sigma_{tT}} \right|_0^\pi \right] 
\rd R \nonumber \\ 
&=& \frac{1}{\sqrt{2\pi \Sigma_{tT}}} \, \re^{-\frac{1}{2} \Sigma_{tT}^{-1} \,\xi_t^2}
\int_0^\infty \re^{  -\frac{1}{2} \Sigma_{tT}^{-1} \,R^2} \left( \re^{R\xi_t/\Sigma_{tT}} - 
\re^{-R\xi_t/\Sigma_{tT}} \right) \rd R \, , 
\label{crypto bond price} 
\end{eqnarray} 
where we have made use of the fact that $\xi_t=1/\pi_t $. We note that the process 
$\{\xi_t\}$ appearing here is the price of the so-called natural numeraire or ``benchmark" 
asset \cite{FH, EP}. Completing the squares in the exponents in (\ref{crypto bond price}), 
we obtain 
\begin{eqnarray}
P_{tT} &=& \frac{1}{\sqrt{2\pi \Sigma_{tT}}} \int_0^\infty 
\left( \re^{  -\frac{1}{2} \Sigma_{tT}^{-1} \,(R-\xi_t)^2} - 
\re^{  -\frac{1}{2} \Sigma_{tT}^{-1} \,(R+\xi_t)^2} \right) \rd R \nonumber \\ &=& 
\frac{1}{\sqrt{\pi}} \int_{-\xi_t/\sqrt{2\Sigma_{tT}}}^{\xi_t/\sqrt{2\Sigma_{tT}}} 
\re^{-u^2} \rd u \, . 
\label{eq:15}
\end{eqnarray} 

\begin{figure}[t!]
\centerline{
\includegraphics[width=0.750\textwidth]{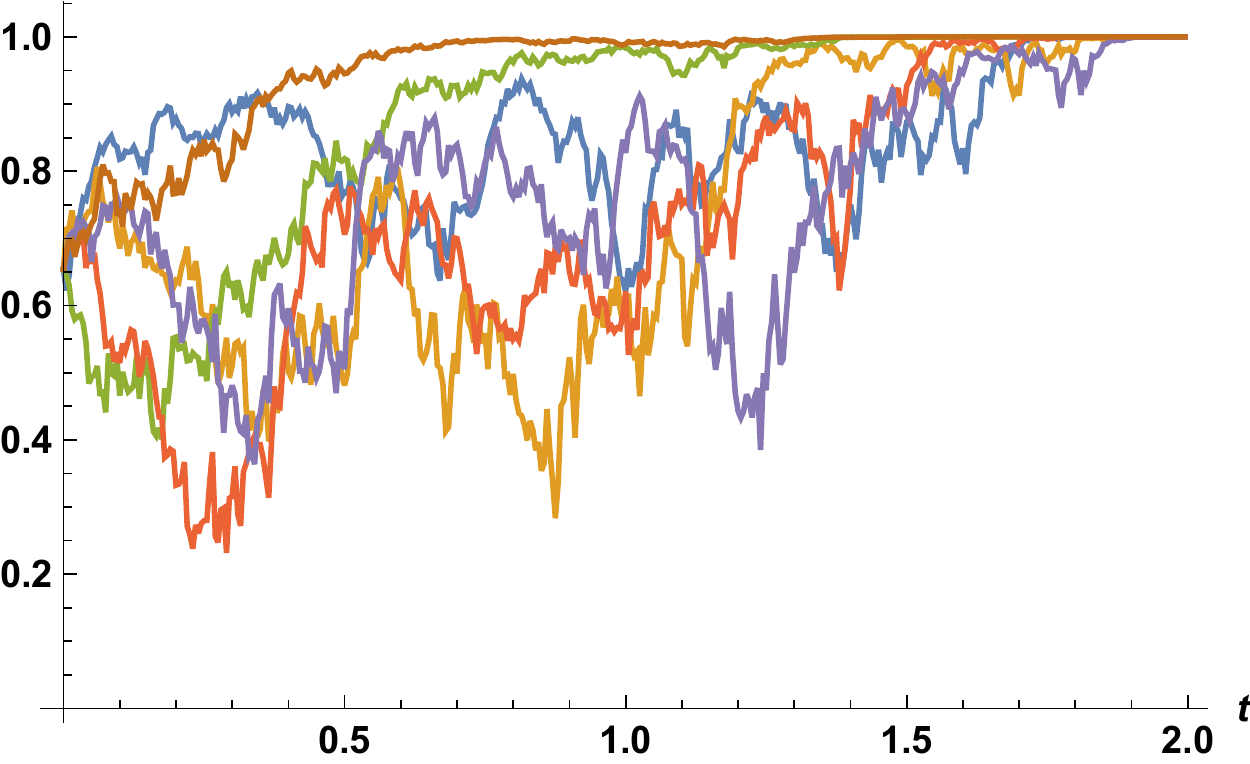}
}
\caption{\footnotesize{
\textit{The discount bond price process $\{P_{tT}\}$ with the choice 
$\sigma_t=0.75$ and $T=2$}. Six sample paths are displayed, 
illustrating the qualitative behaviour of the bond price process in the Bessel(3) 
model when the volatility function $\{\sigma_t\}$ is constant. 
}}
\label{fig:1}
\end{figure}

\noindent Let us now define the error function as usual by setting
\begin{eqnarray}
{\rm erf}(z) = \frac{1}{\sqrt{\pi}} \int_{-z}^z  \re^{-u^2} \rd u 
= \frac{2}{\sqrt{\pi}} \sum_{n=0}^\infty \frac{(-1)^n}{n!(2n+1)}\,z^{2n+1} ,
\label{erf}
\end{eqnarray} 
which is an entire function on the complex plane. Thus, ${\rm erf}(z) = N(\sqrt{2} \, z) - N(-\sqrt{2} \, z)$, where
$N(x)$ is the normal distribution function. 
Then we obtain the following expression for the bond price: 
\begin{eqnarray}
P_{tT} =  {\rm erf} \left( 
\sqrt{\frac{(X_t-a)^2+(Y_t-b)^2+(Z_t-c)^2}{2\,\Sigma_{tT}}} \right).
\label{bond price in terms of erf}
\end{eqnarray} 
Equivalently, we have
\begin{eqnarray}
P_{tT}  = {\rm erf}\left( \frac{\xi_t}{\sqrt{2\Sigma_{tT}}} \right) ,
\end{eqnarray} 
where $\{\xi_t\}$ is the natural numeraire. 
For illustration, we have shown in Figure~\ref{fig:1} some sample paths for the 
bond price. As the price process of an asset, the discount bond 
satisfies the condition that $\{\pi_t P_{tT}\}$ is a martingale, which follows at 
once from the tower property of conditional expectation:
\begin{eqnarray}
{\mathbb E}_s[ \pi_t P_{tT}] = {\mathbb E}_s[ {\mathbb E}_t[ \pi_T]] = {\mathbb E}_s[ \pi_T]
= \pi_s P_{sT}\, . 
\end{eqnarray} 
Alternatively, the martingale condition can be checked directly from 
expression (\ref{bond price in terms of erf}) for the bond price, if we make use of the identity 
\begin{eqnarray}
\frac{1}{\sqrt{\pi\alpha}} \int\limits_0^\infty 
\left( \re^{-\frac{1}{\alpha}(\xi-x)^2} - \re^{-\frac{1}{\alpha}(\xi+x)^2} \right) 
{\rm erf}\left( \frac{\xi}{\sqrt{\beta}}\right)  \rd\xi 
=  {\rm erf}\left( \frac{x}{\sqrt{\alpha+\beta}}\right)\, . 
\end{eqnarray} 

We have seen that the model entails no interest since 
the drift of the pricing kernel is identically zero. Yet, bond prices give rise to 
discounting; in other words, $P_{tT}$ is a decreasing function of $T$ for each $t<T$, since 
$\Sigma_{tT}$ is an increasing of $T$ for $t<T$. One might therefore wonder whether the 
short rate vanishes if one employs the alternative definition of the short rate given by
\begin{eqnarray}
r_t=- \left.\frac {\partial P_{tT} } {\partial T} \right|_{T=t} .
\end{eqnarray}
This can easily be checked. We have 
\begin{eqnarray}
r_t = \left. \frac{\sigma_T^2}{\sqrt{2\pi}\Sigma_{tT}^{3/2}} \, 
\exp\left(-\frac{\xi_t^2}{2\Sigma_{tT}}\right)\right|_{T=t} . 
\end{eqnarray} 
Since $\lim_{t \rightarrow T} \Sigma_{tT} = 0$, the exponential term suppresses the 
right side to give $r_t=0$ for all $t\geq0$. Alternatively, by use of
(\ref{bond price in terms of erf}), a calculation shows that 
\begin{eqnarray}
\rd P_{tT} = \lambda_t \Omega_{tT} P_{tT}\, \rd t + \Omega_{tT}P_{tT} \, \rd W_t , 
\label{bond dynamical equation} 
\end{eqnarray} 
where $\lambda_t =\sigma_t \pi_t$ is the market price of risk, and  where
\begin{eqnarray}
\Omega_{tT} = \frac{2\sigma_t}{P_{tT}\sqrt{2\pi\Sigma_{tT}}} \exp\left( - 
\frac{1}{2\Sigma_{tT}\pi_t^2} \right) 
\end{eqnarray} 
is the discount bond volatility. We note that $\lim_{t \rightarrow T} \Omega_{tT} = 0$. The form of  (\ref{bond dynamical equation}) confirms 
that the contribution $r_t P_{tT}$ normally arising from the short rate in the drift is 
absent. 

\begin{figure}[t!]
\centerline{
\includegraphics[width=0.750\textwidth]{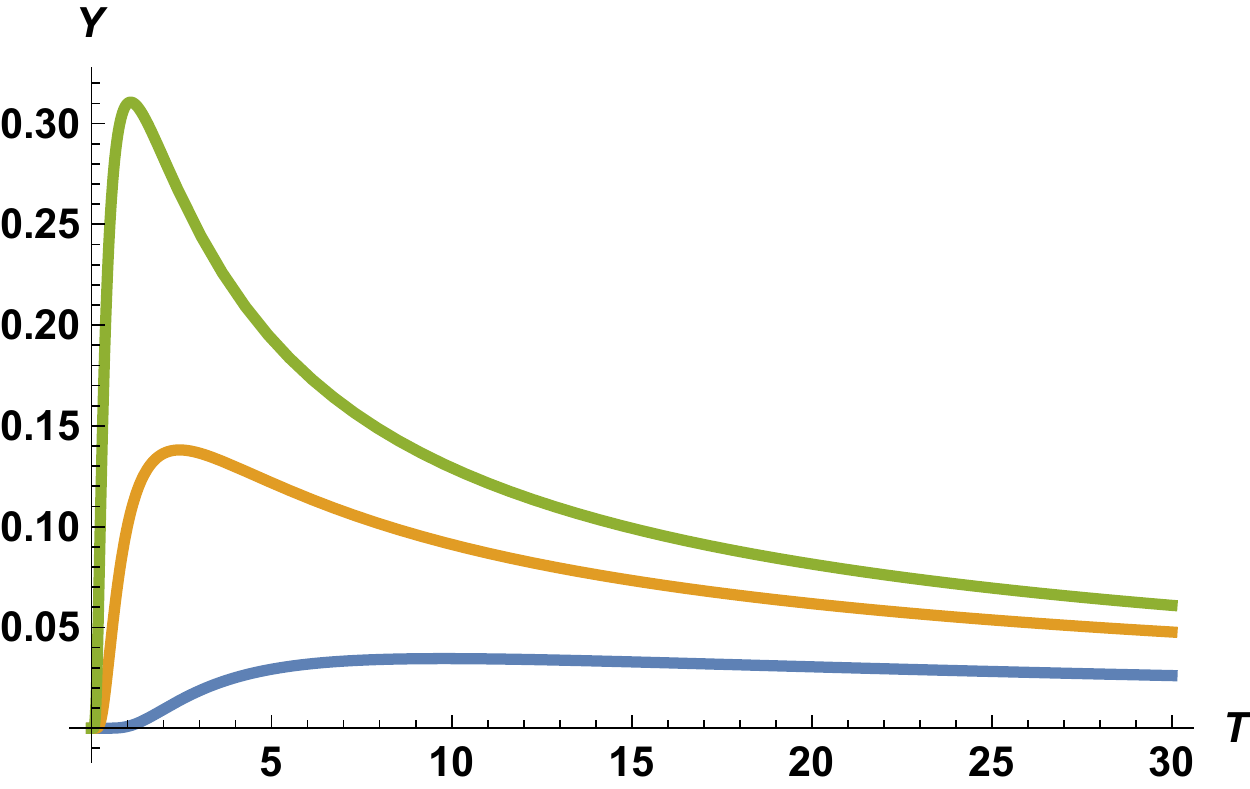}
}
\caption{\footnotesize{
\textit{The initial yield curve ${Y(T)}_{T\geq 0}$ for constant $\sigma_t$}. 
Since $r_0=0$, the yield at $T=0$ vanishes. 
With $\sigma_t=\sigma$, for constant $\sigma$, the 
yield curve peaks and then decays to zero, where 
$Y(T)\sim -\log(\sqrt{2/\pi\sigma T})/T$ as $T\to\infty$. We have 
plotted $\{Y(T)\}$ for $\sigma = 0.3$ (blue), $\sigma =      0.6$ (orange), and 
$\sigma = 0.9$ (green). 
}}
\label{fig:2}
\end{figure}

But the fact that the short rate is zero does not imply that other 
rates are necessarily zero. For instance, as a consequence of the definition
\begin{eqnarray}
f_{tT}=-\frac{\partial\log P_{tT}}{\partial T},
\end{eqnarray} 
a 
calculation shows that the instantaneous forward rates are given by
\begin{eqnarray}
f_{tT} = \frac{\sigma_T^2~\exp\left( - \xi_t^2 / 2\Sigma_{tT}\right)}
{\sqrt{2\pi}\Sigma_{tT}^{3/2}~{\rm erf} \left( \xi_t / \sqrt{2\Sigma_{tT}} \right)}, 
\end{eqnarray} 
and we see that $\lim_{t \rightarrow T} f_{tT} = 0$. Similarly, for the yield curve $\{Y(T)\}$ we obtain the following:
\begin{eqnarray}
Y(T) = -\frac{1}{T} \log\left[ {\rm erf} \left( \frac{1}{\sqrt{2\Sigma_{0T}}} \right) 
\right] .
\end{eqnarray} 
Thus, initial yield curve data can be used to calibrate the freedom in the function $\{\sigma_t\}$. Our calibration scheme is essentially
equivalent to that suggested in \cite{HK} using a time-change technique. 
A typical set of yield curves arising from constant $\{\sigma_t\}$ 
is sketched in Figure~\ref{fig:2}. \\

\noindent \textbf{4. Bond options} \\

\noindent Let us consider the pricing of options on 
discount bonds. To begin, we look at a European-style digital call option with maturity 
$t$ and strike $K$ on a discount bond that matures at time $T$. Thus, the 
option delivers one unit of cryptocurrency at time $t$ in the event that $P_{tT}>K$. 
The option payout is the indicator function
\begin{eqnarray}
H_t = {\mathds 1}\{P_{tT}>K\}, 
\end{eqnarray} 
so the price of a digital call 
is given by
\begin{eqnarray}
D_0 = {\mathbb E}\left[ \frac{1}{\xi_t} {\mathds 1} \!
\left\{ {\rm erf}\left( \frac{\xi_t}{\sqrt{2\Sigma_{tT}}} 
\right) > K \right\} \right] , 
\end{eqnarray} 
where $\xi_t=\pi_t^{-1}$. The error function is increasing in its argument, so we find that there 
is a critical value $\xi^*$ of the natural numeraire such that the option expires in the money if 
$\xi_t>\xi^*$, given by 
\begin{eqnarray}
\xi^* = \sqrt{2\Sigma_{tT}}~{\rm erf}^{-1}(K) . 
\end{eqnarray} 
Therefore, if we switch to a spherical representation for the volume element in 
${\mathds R}^3$, a calculation similar to that presented in (\ref{crypto bond price}) shows 
that the price of a digital call is  
\begin{eqnarray} 
D_0 = 
\frac{1}{2}\left[  {\rm erf}\left( \frac{\xi^*+1}{\sqrt{2\Sigma_{0t}}} 
\right) -  {\rm erf}\left( \frac{\xi^*-1}{\sqrt{2\Sigma_{0t}}} 
\right) \right] .
\end{eqnarray} 

More generally, let us consider the price process $\{D_s\}$ of the digital call 
option, given by the following expression:
\begin{eqnarray} 
D_s=\mathds1_{\{0 \leq s <t\}}\frac{1}{\pi_s}{\mathbb E}_s[\pi_tH_t] \, .
\end{eqnarray} 
Noticing that conditional on ${\mathcal F}_s$ we have 
$X_t\sim{N}(X_s-a,\Sigma_{st})$, $Y_t\sim{N}(Y_s-b,
\Sigma_{st})$ and $Z_t\sim{N}(Z_s-c,\Sigma_{st})$, one finds that a calculation 
analogous to that considered in (\ref{crypto bond price}) leads to the  
formula 
\begin{eqnarray} 
D_s = 
\frac{1}{2\xi_s}\left[  {\rm erf}\left( \frac{\xi^*+\xi_s}{\sqrt{2\Sigma_{st}}} 
\right) -  {\rm erf}\left( \frac{\xi^*-\xi_s}{\sqrt{2\Sigma_{st}}} 
\right) \right] . 
\end{eqnarray} 

We turn now to look at the pricing of an in-arrears caplet, for which the payout at time $T$ is given by
\begin{eqnarray} 
H_T = X( L_{tT}-R)^+ \, ,
\end{eqnarray} 
where $R$ is the cap and $X$ is the notional. 
The crypto rate $L_{tT}$ appearing here is defined by 
\begin{eqnarray}
L_{tT} = \frac{1}{T-t} \left( \frac{1}{P_{tT}} -1 \right) \, . 
\label{Libor rate}
\end{eqnarray}
Since the caplet is paid ``in arrears'', meaning that the payoff is
set at the earlier time $t$ and paid at $T$, and since $L_{tT}$ is known 
at time $t$, we can regard the caplet as a derivative that effectively pays the 
discounted value $H_t=P_{tT}H_T$ at the earlier time $t$. By substitution 
and rearrangement one sees that the effective payout at time 
$t$ takes the form $H_t = N \left( K - P_{tT} \right)^+$, 
where $K$ and $N$ are given by
\begin{eqnarray}
K = \frac{1}{1+R(T-t)} \quad {\rm and} \quad 
N = \frac{X[1+R(T-t)]}{T-t} \, . 
\label{strike and notional}
\end{eqnarray}
Thus we see that a position in an in-arrears caplet is equivalent to a
position in $N$ puts on a discount bond, where the strike
$K$ on the put is the value of a discount bond with simple
crypto yield $R$.
Making use of (\ref{pricing formula}) we deduce that the 
price of 
the caplet is
\begin{eqnarray}
C_0 &=& N \, {\mathbb E}\left[ \pi_t \left(K-P_{tT}\right)^+ \right] \nonumber \\ &=& 
N\, {\mathbb E}\left[ \frac{1}{\xi_t}\left(K-{\rm erf}\left( \frac{\xi_t}{\sqrt{2\Sigma_{tT}}} 
\right)  \right)^+\right] . 
\end{eqnarray} 
If we switch to the spherical representation for the volume element in 
${\mathds R}^3$, a calculation analogous to that presented in (\ref{crypto bond price}) shows 
that the option price can be represented in terms of following Gaussian integrals:
\begin{eqnarray}
C_0 &=& \frac{1}{\sqrt{2\pi\Sigma_{0t}}} \int_0^{\xi^*} \left[ K-{\rm erf}\!\left( 
\frac{R}{\sqrt{2\Sigma_{tT}}}\right) \right] \left( 
\re^{-\frac{1}{2\Sigma_{0t}}(R-1)^2} - 
\re^{  -\frac{1}{2\Sigma_{0t}}(R+1)^2} \right) \rd R \nonumber \\ &=& 
\frac{1}{\sqrt{\pi}} 
\int_{\frac{1}{\sqrt{2\Sigma_{0t}}}}^{\frac{\xi^*+1}{\sqrt{2\Sigma_{tT}}}} 
\re^{-u^2} {\rm erf}\!\left( 
\textstyle{\frac{\sqrt{2\Sigma_{0t}}\,u-1}{\sqrt{2\Sigma_{tT}}}}\right) \rd u - 
\frac{1}{\sqrt{\pi}} 
\int_{\frac{-1}{\sqrt{2\Sigma_{0t}}}}^{\frac{\xi^*-1}{\sqrt{2\Sigma_{tT}}}} 
\re^{-u^2} {\rm erf}\!\left( 
\textstyle{\frac{\sqrt{2\Sigma_{0t}}\,u+1}{\sqrt{2\Sigma_{tT}}}}\right) \rd u
\nonumber \\ && +K \left[ {\rm erf}\!\left( 
\textstyle{\frac{1}{\sqrt{2\Sigma_{0t}}}} \right)
- \half \left( {\rm erf}\!\left( 
\textstyle{\frac{\xi^*+1}{\sqrt{2\Sigma_{0t}}}} \right)- {\rm erf}\!\left( 
\textstyle{\frac{\xi^*-1}{\sqrt{2\Sigma_{0t}}}} \right) \right) \right]. 
\end{eqnarray} 
While there appears to be no simpler representation for the Gaussian 
integrals appearing here, numerical evaluation is 
straightforward. \\

\noindent \textbf{5. Models based on higher-order Bessel processes} \\

\noindent Bessel processes of order four or more also give rise to cryptobond models, 
 analogous to the one we have already investigated. In particular, if we 
consider a collection of Gaussian processes $\{X_t^a\}$ for $a=1,2,\ldots,n$ of the type given by 
(\ref{Brownian motions}) in dimension $n\geq3$, then we can model the pricing kernel 
by setting  
\begin{eqnarray} 
\pi_t=\left[(X_t^1)^2+\cdots+(X_t^n)^2\right]^{(2-n)/2} . 
\end{eqnarray} 
A short calculation shows that 
\begin{eqnarray} 
\rd\pi_t = -(n-2)\, \sigma_t \, \pi_t^{(n-1)/(n-2)} \rd W_t \,,
\label{eq:39}
\end{eqnarray} 
from which it follows on 
account of the test discussed in \cite{BE,MU} that $\{\pi_t\}$ is a strict local martingale for all $n\geq3$ . 

As an illustration, we present a pricing kernel model for 
cryptocurrencies based on the reciprocal of the Bessel process in four 
dimensions. See, for example, reference  \cite{EP} for properties of the Bessel(4) process.  
For our model we take 
\begin{eqnarray}
\pi_t=\frac{1}{(X_t^1-a)^2+(X_t^2-b)^2+(X_t^3-c)^2+(X_t^4-d)^2} \,,
\end{eqnarray} 
where the $\{X_t^k\}_{k=1,\ldots,4}$ are four independent Gaussian processes of 
the form (\ref{Brownian motions}), and the constants $a,b,c,d$ are chosen such that $a^2+b^2+c^2+d^2=1$. A calculation 
shows that the dynamical equation of the pricing kernel is
\begin{eqnarray}
\rd\pi_t = -2 \sigma_t \, \pi_t^{3/2} \rd W_t \, ,
\end{eqnarray} 
which corresponds to 
(\ref{eq:39}) for $n=4$. We wish to compute the discount bond price in this 
model, which, following the logic of (\ref{eq:15.0}), with 
${\boldsymbol\xi}_t=(X_t^1-a,X_t^2-b,X_t^3-c,X_t^4-d)$, is given by
\begin{eqnarray}
P_{tT} = \frac{1}{\pi_t} 
\frac{1}{(\sqrt{2\pi \Sigma_{tT}})^4} \int_{{\mathds R}^4} 
\frac{1}{R^2}\, \re^{-\frac{1}{2} \Sigma_{tT}^{-1} \,|{\boldsymbol R}-{\boldsymbol\xi}_t|^2} 
\rd^4 {\boldsymbol R} \, .
\end{eqnarray} 
We switch to a spherical representation. In four dimensions, we set 
$x=R\sin\theta \, \sin\varphi\, \cos\phi$, 
$y=R\sin\theta\, \sin\varphi\, \sin\phi$,   
$z=R\sin\theta \, \cos\varphi$, and   
$w=R\cos\theta$, with the volume element 
\begin{eqnarray}
\rd^4 {\boldsymbol R}=R^3 \sin^2\theta\, \sin\varphi\, \rd R \,\rd\theta\, 
\rd\varphi\, \rd\phi \,. 
\end{eqnarray} 
Note that $\theta,\varphi\in[0,\pi]$ and $\phi\in[0,2\pi]$. 
Since the vector ${\boldsymbol\xi}_t$ is fixed, and because of the spherical 
symmetry, we may without loss of generality choose ${\boldsymbol\xi}_t$ to be in the direction 
of the $w$-axis. A similar assumption was made in the three-dimensional case in 
(\ref{crypto bond price}), where ${\boldsymbol\xi}_t$ was taken to be in the 
$z$-direction. Then we have ${\boldsymbol R}\cdot{\boldsymbol\xi}_t = 
R \xi_t \cos\theta$,  a choice that simplifies the calculation somewhat. 
Integration over $\phi$ gives $2\pi$, whereas $\int_0^\pi \sin\varphi \, 
\rd \varphi=2$, so after performing the integration over these variables we obtain 
\begin{eqnarray}
P_{tT} = \frac{1}{\pi_t} 
\frac{2}{2\pi \Sigma_{tT}^{2}} 
\int_0^\infty  R  
\int_0^\pi \sin^2\theta\, \re^{ -\frac{1}{2} \Sigma_{tT}^{-1} \,
(R^2-2R\xi_t\cos\theta+\xi_t^2)} \rd\theta \, \rd R \,,
\end{eqnarray} 
where $\{\xi_t\}$ represents the Bessel process in four dimensions, so 
$\pi_t=\xi_t^{-2}$. To proceed, we note the identity 
\begin{eqnarray}
\int_0^\pi \sin^2\theta \, \re^{\nu\cos\theta} \rd\theta = \frac{\pi}{\nu}\, I_1(\nu) . 
\end{eqnarray} 
This follows if we observe that $\sin^2\theta \, \re^{\nu\cos\theta} = (\sin\theta) 
(\sin\theta \, \re^{\nu\cos\theta})$, and that $\sin\theta \, \re^{\nu\cos\theta} = 
-\nu^{-1} \partial_\theta \, \re^{\nu\cos\theta}$, which shows that we can perform 
integration by parts to reduce the integrand to $\cos\theta \, \re^{\nu\cos\theta}$. 
But then we 
notice that $\cos\theta \, \re^{\nu\cos\theta}=\partial_\nu\,\re^{\nu\cos\theta}$, 
so moving $\partial_\nu$ outside the integration we see that the integrand reduces further 
to $\re^{\nu\cos\theta}$. But this gives rise to a Bessel function, and we have 
$\int_0^\pi \re^{\nu\cos\theta} \rd\theta = \pi \, I_0(\nu)$. 
Differentiating and using the differential identity $\partial_\nu\, I_0(\nu) 
= I_1(\nu)$, we arrive at the conclusion. 
Alternatively, if we recall the definition 
\begin{eqnarray}
I_n(\nu) = \frac{1}{\pi} \int_0^\pi \re^{\nu\cos\theta} \cos(n\theta) \rd \theta
\end{eqnarray} 
for the generalized Bessel function of the first kind, we arrive at the same 
conclusion more expediently. In any case, we deduce that 
\begin{eqnarray}
\int_0^\pi \sin^2\theta\, \re^{ R \, \xi_t \, \Sigma_{tT}^{-1} \,
\cos\theta} \rd\theta = \frac{\pi\Sigma_{tT}}{R\xi_t} \, I_1\!\left( 
{\textstyle \frac{R\xi_t}{\Sigma_{tT}} } \right) . 
\end{eqnarray} 
Thus using $(\pi_t\xi_t)^{-1}=\xi_t$ we obtain 
\begin{eqnarray}
P_{tT} = \frac{\xi_t}{\Sigma_{tT}} 
\int_0^\infty \re^{ -\frac{1}{2} \Sigma_{tT}^{-1} \,
(R^2+\xi_t^2)} I_1\!\left( 
{\textstyle \frac{R\xi_t}{\Sigma_{tT}} } \right)
\, \rd R . 
\end{eqnarray} 
If we define $u=R/\sqrt{\Sigma_{tT}}$ and $\eta_t=\xi_t/\sqrt{\Sigma_{tT}}$, the 
expression simplifies to 
\begin{eqnarray}
P_{tT} = \eta_t \, \re^{ -\frac{1}{2}\eta_t^2} 
\int_0^\infty \re^{ -\frac{1}{2}u^2} I_1\!\left(\eta_t u\right) 
\, \rd u . 
\end{eqnarray} 
Now we use the identity 
\begin{eqnarray}
\int_0^\infty \re^{-\frac{1}{2}u^2} I_1(\eta u)\, \rd u = 
\frac{\re^{\frac{1}{2}\eta^2} - 1}{\eta} , 
\label{eq:49} 
\end{eqnarray} 
which can be established by use of the Taylor series expansion of the 
Bessel function 
\begin{eqnarray}
I_n(\nu) =  \sum_{k=0}^\infty \frac{1}{2^{2k+n} k! 
\Gamma(n+k+1)}\, \nu^{2k+n} 
\label{eq:50} 
\end{eqnarray} 
along with the expression 
\begin{eqnarray}
\int_0^\infty \re^{-\frac{1}{2}u^2} u^{2k+n} \rd u = 2^{(2k+n-1)/2} \, 
\Gamma\left(\frac{2k+n+1}{2}\right) 
\label{eq:51} 
\end{eqnarray} 
for the Gaussian moments. Specifically, substituting (\ref{eq:50}) for $n=1$ and 
$\nu=\eta u$ in the left side of (\ref{eq:49}) and using (\ref{eq:51}) for $n=1$, we 
obtain 
\begin{eqnarray}
\int_0^\infty \re^{-\frac{1}{2}u^2} I_1(\eta u)\, \rd u &=& 
\sum_{k=0}^\infty \frac{\eta^{2k+1}}{2^{2k+1}k!\Gamma(k+2)} \, 2^k 
\Gamma(k+1)  \nonumber \\ &=& 
\sum_{k=0}^\infty \frac{(\eta/2)^{k+1}}{(k+1)!} \nonumber \\ &=& 
\frac{1}{\eta} \left( \sum_{k=0}^\infty \frac{(\eta/2)^k}{k!} - 1 \right) , 
\label{eq:52} 
\end{eqnarray} 
and this establishes (\ref{eq:49}). 
Putting these together we arrive at the bond price 
\begin{eqnarray} 
P_{tT} = 1 - \re^{ -\frac{1}{2} \Sigma_{tT}^{-1} \,\xi_t^2} , 
\label{eq:41} 
\end{eqnarray}
which turns out to be surprisingly simple. 

\begin{figure}[t!]
\centerline{
\includegraphics[width=0.750\textwidth]{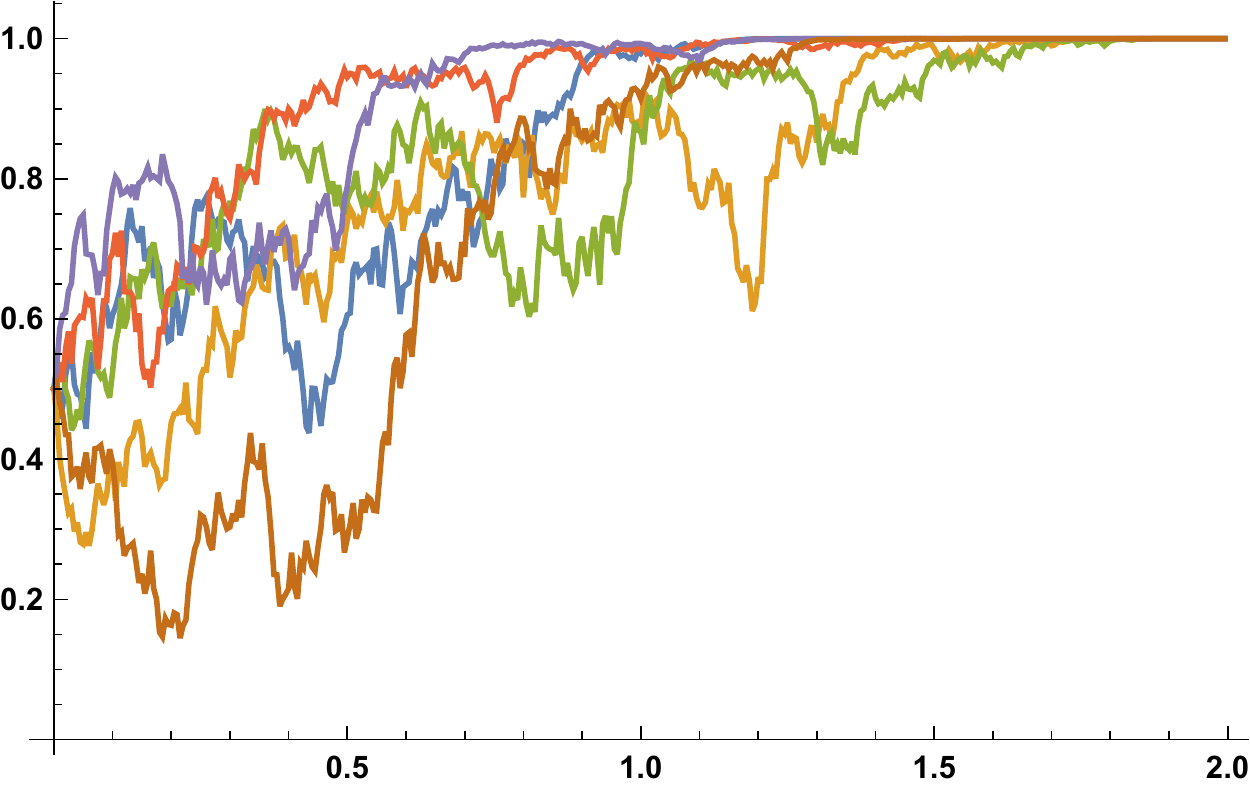}
}
\caption{\footnotesize{
\textit{The discount bond price process with 
$\sigma_t=0.6$ and $T=2$}. Six sample paths are displayed, 
illustrating the qualitative behaviour of the bond price in a 
model based on the Bessel(4) process. 
}}
\label{fig:3}
\end{figure}

For illustration we have shown in 
Figure~\ref{fig:3} some sample paths for the bond price process. 
Note that  $\lim_{t\to T}P_{tT} =1$; whereas, 
assuming that $\sigma_t>0$ for all $t\geq0$, 
we have $\lim_{T\to\infty}\Sigma_{tT}=\infty$, from which it follows that $\lim_{T\to\infty} P_{tT}\to0$. The initial 
bond price is given by 
\begin{eqnarray} 
P_{0T}=1-\exp{ \left(-\frac{1}{2\Sigma_{0T}}\right)},  
\end{eqnarray}
from 
which we deduce that the initial yield curve takes the form
\begin{eqnarray} 
Y(T) = -\frac{1}{T} \log\left[1-\exp{ \left(-\frac{1}{2\Sigma_{0T}}\right)} \right] .
\end{eqnarray} 
This relation can be used to calibrate the volatility function to market data. Specifically, we have 
\begin{eqnarray}
\sigma_T^2 = - \frac{\left(Y(T)+T Y'(T)\right)\re^{-TY(T)}}{2(1-
\re^{-T Y(T)})\left(\log(1-\re^{-TY(T)})\right)^2} \, ,
\end{eqnarray}
which allows us to determine the form of the function $\{\sigma_t\}_{t\geq 0}$ from any initial yield curve $\{Y(t)\}_{t\geq0}$ satisfying the constraint $Y(0)=0$. 

Next we examine the dynamics of the bond price. If we start with  
\begin{eqnarray}
\rd\xi_t = \frac{3\sigma_t^2}{2\xi_t}\,\rd t + \sigma_t \, \rd W_t \, ,
\end{eqnarray}
an application of Ito's formula gives 
\begin{eqnarray}
\rd P_{tT} = \lambda_t \Omega_{tT} P_{tT}\, \rd t + \Omega_{tT}P_{tT} \, \rd W_t  \, ,
 \end{eqnarray}
where 
\begin{eqnarray}
\lambda_t = 2 \sigma_t \xi_t^{-1} \qquad {\rm and} \qquad 
\Omega_{tT} = \frac{\sigma_t \xi_t}{P_{tT} \Sigma_{tT}} \, 
\re^{ -\frac{1}{2} \Sigma_{tT}^{-1} \,\xi_t^2} .
\end{eqnarray}
This result offers an independent confirmation of the fact that $r_t=0$ for all 
$t\geq0$ in the present model. 

Let us now consider the valuation of a call option on a discount bond. The payoff takes the form
$H_t = (P_{tT}-K)^+$, where $K$ is the strike price of the option, $t$ is the expiration date of the option, and $T> t$ is the maturity date of the bond. We assume that $0 < K < 1$. Since the bond price is an 
increasing function of $\xi_t$, we find that there is a critical value 
\begin{eqnarray}
\xi^* = \sqrt{-2\Sigma_{tT}\log(1-K)} 
\end{eqnarray} 
such that $H_t=0$ if $\xi_t\leq\xi^*$. After we perform the integration 
over the $(\phi,\varphi)$ variables, we find that the initial price of the option is determined 
by the integral 
\begin{eqnarray}
C_{0} = \frac{1}{\pi \Sigma_{0t}^{2}} 
\int\limits_{\xi^*}^\infty R \left( (1-K)-\re^{-\frac{1}{2\Sigma_{tT}}R^2} \right)  
\re^{-\frac{1}{2\Sigma_{0t}}(R^2+1)}
\int\limits_0^\pi \sin^2\theta\, \re^{ \frac{R}{\Sigma_{0t}}
\cos\theta} \rd\theta \, \rd R \, . 
\end{eqnarray} 
Performing the $\theta$ integration, we thus have 
\begin{eqnarray}
C_{0} = \frac{1}{\Sigma_{0t}} 
\int_{\xi^*}^\infty  \left( (1-K)-\re^{-\frac{1}{2\Sigma_{tT}}R^2} \right)  
I_1\left(\frac{R}{\Sigma_{0t}}\right) \re^{-\frac{1}{2\Sigma_{0t}}(R^2+1)}
 \, \rd R \, . 
 \label{eq:61} 
\end{eqnarray} 
Similarly, for a put option with payout $(K-P_{tT})^+$, we obtain 
\begin{eqnarray}
P_{0} = \frac{1}{\Sigma_{0t}} 
\int_0^{\xi^*}  \left( (K-1)+\re^{-\frac{1}{2\Sigma_{tT}}R^2} \right)  
I_1\left(\frac{R}{\Sigma_{0t}}\right) \re^{-\frac{1}{2\Sigma_{0t}}(R^2+1)}
 \, \rd R \, ,
 \label{eq:61.2} 
\end{eqnarray} 
from which we observe that 
\begin{eqnarray}
C_0 - P_{0} = \frac{1}{\Sigma_{0t}} 
\int_0^{\infty}  \left( (K-1)+\re^{-\frac{1}{2\Sigma_{tT}}R^2} \right)  
I_1\left(\frac{R}{\Sigma_{0t}}\right) \re^{-\frac{1}{2\Sigma_{0t}}(R^2+1)}
 \, \rd R \, .
 \label{eq:61.3} 
\end{eqnarray} 
Using (\ref{eq:49}) we can integrate the right side of (\ref{eq:61.3}) 
explicitly to obtain the put-call parity relation: 
\begin{eqnarray}
C_0-P_0 &=& (1-K)\left(1-\re^{-\frac{1}{2\Sigma_{0t}}} \right)+
\re^{-\frac{1}{2\Sigma_{0t}}} - 
\re^{-\frac{1}{2\Sigma_{0t}}\left[ 1-\frac{\Sigma_{tT}}{\Sigma_{0T}}\right]} 
\nonumber \\ &=& P_{0T}-KP_{0t} ,
\end{eqnarray} 
where we have made use of the fact that $\Sigma_{0t}+\Sigma_{tT}=
\Sigma_{0T}$. 

The indefinite Gaussian integrals of the Bessel function for the option 
prices have to be evaluated 
numerically. It is interesting to note that despite the simplicity in the model for the 
bond price, the option price cannot be expressed in closed form in terms of 
known functions. Nevertheless, fast numerical valuation is straightforward. 
To see this, we use the Taylor series expansion (\ref{eq:50}) for the Bessel 
function to obtain 
\begin{eqnarray}
C_0 =  \frac{1}{\Sigma_{0t}}  \re^{-\frac{1}{2\Sigma_{0t}}}\sum_{k=0}^\infty 
\frac{(1/2\Sigma_{0t})^{2k+1}}{k! (k+1)!} \int_{\xi^*}^\infty R^{2k+1} \left[ 
(1-K)-\re^{-\frac{1}{2\Sigma_{tT}}R^2} \right]
\re^{-\frac{1}{2\Sigma_{0t}}R^2} \rd R \, . 
\label{eq:69a} 
 \end{eqnarray}
Then, by changing the integration variable by setting $u=R^2$, we find that the 
integration reduces to that of an incomplete gamma function
\begin{eqnarray}
\Gamma(a,z) = \int_z^\infty u^{a-1} \re^{-u} \rd u , 
\label{eq:63} 
\end{eqnarray} 
and we thus obtain the option price in the form of a series: 
\begin{eqnarray}
C_0 &=&  \re^{-\frac{1}{2\Sigma_{0t}}} \sum_{k=0}^\infty 
\frac{(1/2\Sigma_{0t})^{k+1}}{k! (k+1)!} \left[ (1-K) 
\, \Gamma\left( k+1,-\frac{\Sigma_{tT}}{\Sigma_{0t}}\log(1-K) \right) \right. 
\nonumber \\ 
&& \qquad \qquad \qquad \qquad \qquad \left. - 
\left( \frac{\Sigma_{tT}}{\Sigma_{0T}}\right)^{k+1} 
\Gamma\left( k+1,-\frac{\Sigma_{0T}}{\Sigma_{0t}}\log(1-K) \right)
\right]. 
\end{eqnarray}
On account  of the appearance of the double factorial in the denominator in the 
summand in the expression above, the series converges rapidly, making it a useful expression for 
numerical valuation of the option price. 

In particular, truncating the sum at, 
say, $k=20$, we can obtain option prices very rapidly by use of standard numerical tools; 
the difference of the result thus obtained and the result of a standard numerical valuation of the integral (\ref{eq:69a}) 
is of the order $10^{-16}$. 

More generally, in higher dimensions 
it should be evident that by use of the spherical representation for calculating the 
expectation ${\mathbb E}_t[\pi_T]$ it is always possible to set the direction of 
the vector ${\boldsymbol\xi}_t$ such that 
${\boldsymbol R}\cdot{\boldsymbol\xi}_t = R \xi_t \cos\theta$. Thus the only 
nontrivial integration concerns the variables $\theta$ and $R$. Performing the 
$\theta$ integration we arrive at a linear combination of Bessel functions if the 
dimension $n$ of the Bessel process is even, which then has to be integrated 
with respect to a Gaussian measure to obtain an expression for the bond price; 
whereas if $n$ is odd, the $\theta$ integration gives rise to a linear 
combination of exponential functions, which again has to be integrated 
with respect to a Gaussian measure to obtain an expression for the bond price. 
Thus, 
depending on whether $n$ is even or odd, for $n\geq3$ we obtain two different types 
of cryptocurrency interest rate models. \\ 

\noindent \textbf{6. Complex extensions of the model}  \\

\noindent 
The model associated with the reciprocal of the Bessel process in three 
dimensions can be extended in an alternative manner to allow for 
parametric degrees of freedom to be incorporated. 
This can be achieved if we allow the parameters $a$, $b$, and $c$ appearing in 
(\ref{pi}) to be complex numbers.  The real part of  
the resulting complexified process  
$\{\pi_t^{\mathds C}\}$ then defines an admissible model for the pricing kernel, with 
vanishing short rate. The reason for this is that when the parameters $a$, $b$, 
and $c$ are complex, then both real and imaginary parts of the function 
$u(x,y,z)$ satisfy Laplace's equation, and the real part is strictly positive.  The additional freedom thus arising can be used, for instance, to calibrate the model not only against the yield curve but also against option 
prices. 

To proceed, let us therefore write $a=a_0+\ri a_1$, $b=b_0+\ri b_1$ and $c=c_0+\ri c_1$. 
Additionally, let us write ${\tilde x}=x-a_0$, ${\tilde y}=y-b_0$ and ${\tilde z}=
z-c_0$. Then we have 
\begin{eqnarray}
u(x,y,z) = \sqrt{
\frac{{\tilde x}^2+{\tilde y}^2+{\tilde z}^2-a_1^2-b_1^2-c_1^2+2{\rm i} 
({\tilde x}a_1+{\tilde y}b_1+{\tilde z}c_1)}
{\left({\tilde x}^2+{\tilde y}^2+{\tilde z}^2-a_1^2-b_1^2-c_1^2\right)^2 + 4\left( 
{\tilde x}a_1+{\tilde y}b_1+{\tilde z}c_1\right)^2} 
} \,  .
\end{eqnarray}
This function is a solution of Laplace's equation away from 
the ``ring" singularity defined by the intersection of the two-sphere $\Sigma$ of radius 
$\sqrt{a_1^2+b_1^2+c_1^2}$ centred at the point $(a_0,b_0,c_0)$ and the 
two-plane $\Pi$ defined by $a_1 x + b_1 y + c_1 z = a_0a_1+b_0b_1+c_0c_1$ which passes through the point $(a_0,b_0,c_0)$ and hence cuts $\Sigma$ in an equatorial circle. 
We recall the formula 
\begin{eqnarray}
\sqrt{A+\ri B} = \sqrt{\frac{A+\sqrt{A^2+B^2}}{2}} + \ri \frac{B}{|B|} 
\sqrt{\frac{-A+\sqrt{A^2+B^2}}{2}}  
\end{eqnarray}
for the real and the imaginary parts of the principal square-root of a complex number for which $B \neq 0$. 
It follows that ${\rm Re} \, (u )> 0$ on ${\mathds R}^3 \backslash \{\Sigma \cap \Pi \}$, whereas ${\rm Im} \, (u )= 0$ on $\Pi \backslash \{\Sigma \cap \Pi \}$.  With these results at hand, we introduce a new crypto-rate model by 
setting
\begin{eqnarray}
\pi_t = {\rm Re} \, \left(\pi_t^{\mathds C}\right).
\end{eqnarray}
Then writing 
${\tilde X}_t=X_t-a_0$, ${\tilde Y}_t=Y_t-b_0$, and ${\tilde Z}_t=Z_t-c_0$, we 
obtain the following expression for the pricing kernel: 
\begin{eqnarray} 
\pi_t = \sqrt{\frac{
{{\tilde X}_t^2+{\tilde Y}_t^2+{\tilde Z}_t^2 \atop \quad 
-a_1^2-b_1^2-c_1^2}+
\sqrt{
{\left({\tilde X}_t^2+{\tilde Y}_t^2+{\tilde Z}_t^2-a_1^2-b_1^2-c_1^2\right)^2 
\atop \quad 
+ 4\left( {\tilde X}_t a_1+{\tilde Y}_t b_1+{\tilde Z}_t c_1\right)^2}}}
{2\left[\left({\tilde X}_t^2+{\tilde Y}_t^2+{\tilde Z}_t^2-a_1^2-b_1^2-c_1^2\right)^2 
+ 4\left( {\tilde X}_t a_1+{\tilde Y}_t b_1+{\tilde Z}_t c_1\right)^2\right]}} \, . \, \, 
\label{eq:69}
\end{eqnarray}
The normalization $\pi_0=1$  imposes one constraint, whereas rotational 
symmetry can be used to eliminate two further parameters. Thus we are left with 
a model with three exogenously specifiable parameters that 
can be used to fit option prices. 

To obtain an expression for the bond price, we need to work out the conditional 
expectation ${\mathbb E}_t[\pi_T]$. Rather than using expression 
(\ref{eq:69}) for the pricing kernel, which makes the computation somewhat 
cumbersome, we can take advantage of the fact that 
\begin{eqnarray} 
{\mathbb E}_t \left[ {\rm Re} 
\left(\pi_t^{\mathds C}\right)\right] = 
{\rm Re}\left({\mathbb E}_t \left[\pi_T^{\mathds C}\right]\right). 
\end{eqnarray}
Then we can 
use the simpler formula (\ref{pi}) for the pricing kernel, with complex parameters 
$a$, $b$, and $c$, and calculate its expectation, taking the real part of 
the result. Using the spherical representation, we have 
\begin{eqnarray}
{\mathbb E}_t[\pi_T^{\mathds C}] = 
\frac{1}{(\sqrt{2\pi \Sigma_{tT}})^3} \int_{{\mathds R}^3} 
\frac{1}{R}\, \re^{-\frac{1}{2\Sigma_{tT}}|{\boldsymbol R}-({\boldsymbol\xi_t}
-{\rm i}{\boldsymbol\delta})|^2} 
R^2 \sin\theta\, \rd R \,\rd\theta\, \rd\phi , 
\end{eqnarray} 
where ${\boldsymbol\xi_t}=(X_t-a_0,Y_t-b_0,Z_t-c_0)$ and 
${\boldsymbol\delta}=(a_1,b_1,c_1)$. 

It turns out that the calculation leading to 
(\ref{bond price in terms of erf}) is applicable for complex parameters 
$a$, $b$, and $c$. 
To see this, we perform the integration explicitly. Recall that 
in the real case 
where ${\boldsymbol\delta}={\boldsymbol0}$ we have one fixed vector 
${\boldsymbol\xi_t}$ in the exponent of the integrand, so by using the spherical 
symmetry we 
choose this vector to point in the $z$ direction, resulting in the simple expression 
${\boldsymbol R}\cdot{\boldsymbol\xi_t}=R\xi_t\cos\theta$, which was used 
in the calculation of (\ref{bond price in terms of erf}). 

In the present case, we 
have two fixed vectors ${\boldsymbol\xi_t}$ and ${\boldsymbol\delta}$, so we can use the rotational symmetry to 
let the two vectors lie on the 
$x$-$y$ plane, symmetrically placed about the $x$-axis. We let $2\alpha_t$ 
denote the angle between the two vectors ${\boldsymbol\xi_t}$ and 
${\boldsymbol\delta}$. In other words, we have  ${\boldsymbol\xi_t}\cdot{\boldsymbol\delta}=\xi_t \,
\delta \cos(2\alpha_t)$, where $\xi_t^2={\boldsymbol\xi_t}\cdot{\boldsymbol\xi_t}$ 
and $\delta^2={\boldsymbol\delta}\cdot{\boldsymbol\delta}$. Thus, the angle 
between ${\boldsymbol\xi_t}$ and the $x$-axis is $\alpha_t$, and similarly the 
angle between ${\boldsymbol\delta}$ and the $x$-axis is $-\alpha_t$. With this 
choice of coordinates we have 
\begin{eqnarray} 
{\boldsymbol R}\cdot ({\boldsymbol\xi_t}-{\rm i}{\boldsymbol\delta}) = 
\big( R(\xi_t-\ri\delta)\sin\theta\cos\alpha_t\big)\cos\phi 
+  \big( R(\xi_t+\ri\delta)\sin\theta\sin\alpha_t\big)\sin\phi . 
\end{eqnarray} 
We are now in a position to perform the integration over the variable $\phi$. 
To this end we recall the identity 
\begin{eqnarray} 
\int_0^{2\pi} \re^{p\cos\phi+q\sin\phi} \rd \phi = 2\pi I_0\left( \sqrt{p^2+q^2}\right) . 
\end{eqnarray} 
This can be seen by viewing the exponent of the integrand as an inner product 
between the vector $(p,q)$ and the unit vector placed at an angle $\phi$ from 
the vector $(p,q)$. Then the exponent is equivalent to $ \sqrt{p^2+q^2}\cos\phi$, 
and the 
result follows. In the present case we have $p=R(\xi_t-\ri\delta)\sin\theta\cos\alpha_t$ and $q=R(\xi_t+\ri\delta)\sin\theta\sin\alpha_t$, so  
$p^2+q^2=R^2\omega_t^2\sin^2\theta$, where 
\begin{eqnarray}
\omega_t^2 = |({\boldsymbol\xi_t}-{\rm i}{\boldsymbol\delta})|^2 = 
\xi_t^2-\delta^2-2\ri\xi_t\delta\cos(2\alpha_t) .
\label{eq:73} 
\end{eqnarray}
We thus deduce that 
\begin{eqnarray} 
{\mathbb E}_t\left[\pi_T^{\mathds C}\right] = \frac{2\pi}{(\sqrt{2\pi \Sigma_{tT}})^3} 
\int_0^\infty \int_0^\pi R\, \re^{-\frac{1}{2\Sigma_{tT}}(R^2+\omega_t^2)} \sin\theta 
I_0\left( \frac{R\omega_t}{\Sigma_{tT}}\sin\theta \right) \rd\theta \rd R .
\end{eqnarray} 
To perform the integration over the variable $\theta$ we note from (\ref{eq:50}) 
that 
\begin{eqnarray} 
I_0(\nu\sin\theta)=\sum_{k=0}^\infty \frac{\nu^{2k}}{2^{2k}(k!)^2~}(\sin\theta)^{2k} . 
\end{eqnarray} 
Thus, because
\begin{eqnarray} 
\int_0^\pi (\sin\theta)^{2k+1} \rd \theta = \frac{2^{2k+1}(k!)^2}{(2k+1)!} \,, 
\end{eqnarray} 
and taking into account the Taylor series expansion 
\begin{eqnarray}
\frac{2\sinh(\nu)}{\nu}=2\sum_{k=0}^\infty \frac{\nu^{2k}}{(2k+1)!}\, ,
\end{eqnarray} 
we deduce the identity 
\begin{eqnarray} 
\int_0^\pi \sin\theta \, I_0(\nu\sin\theta) \rd \theta = \frac{1}{\nu}\left( \re^{\nu} 
- \re^{-\nu}\right) ,
\end{eqnarray} 
from which it follows that 
\begin{eqnarray} 
{\mathbb E}_t\left[\pi_T^{\mathds C}\right] &=& 
\frac{\omega_t^{-1}}{\sqrt{2\pi \Sigma_{tT}}} 
\int_0^\infty \re^{-\frac{1}{2\Sigma_{tT}}(R^2+\omega_t^2)} \left( 
\re^{\frac{R\omega_t}{\Sigma_{tT}}} - \re^{-\frac{R\omega_t}{\Sigma_{tT}}} 
\right) \rd R \nonumber \\ &=& 
\omega_t^{-1} \, {\rm erf}\left( \frac{\omega_t}{\sqrt{2\Sigma_{tT}}}\right) . 
\end{eqnarray} 
Noting that $\pi_t^{\mathds C}=\omega_t^{-1}$ we thus validate the 
claim that the calculation leading to (\ref{bond price in terms of erf}) is 
applicable for complex parameters $a$, $b$, and $c$. In particular, for 
the bond price we have
\begin{eqnarray} 
P_{tT} =\frac{1}{{\rm Re}(\omega_t^{-1})} \, {\rm Re}\left( 
\omega_t^{-1} \, {\rm erf}\left( \frac{\omega_t}{\sqrt{2\Sigma_{tT}}}\right) \right) , 
\label{eq:77} 
\end{eqnarray} 
where $\omega_t$ is defined by (\ref{eq:73}). 
It should be apparent 
that in the real case for which ${\boldsymbol\delta}={\boldsymbol0}$, we recover 
from (\ref{eq:77}) the previous expression (\ref{bond price in terms of erf}) for the 
bond price. 

Hence, by the complexification of models based on Bessel processes we can obtain genuine parametric extensions 
of the resulting term structure models.  The complexification method that we have applied here is reminiscent of an analogous technique that has been used in physical applications \cite{Newman, Synge}. \\

\noindent \textbf{7. Discussion}  \\

\noindent The notion that strict local martingales should play a role in finance has been considered 
in various contexts by a number of authors.  One can mention, in particular, the so-called 
benchmark approach of Platen and his collaborators, and the F\"ollmer-Jarrow-Protter theory of price 
bubbles (see \cite{EP,PJ,Jarrow} and references cited therein) as examples that have attracted considerable 
attention. In the present paper, we have put forward an altogether different proposal for the application 
of local martingales in the theory of finance --- namely, the idea that strict local martingales can be 
used as a basis for modelling the pricing kernel in a crypto economy where there is no money market 
account. The familiar rules of risk-neutral pricing no longer apply, since the money market account is 
not available to act as a numeraire. Nevertheless, as we have shown, the existence of a market price 
of risk is sufficient to ensure nontrivial discounting, despite the vanishing of the short rate.  Our 
approach to crypto interest rates has been developed in some detail in models based on Bessel 
processes of order three and order four. These models have the advantage that explicit formulae, or 
semi-explicit expressions involving Gaussian integrals, can be obtained for the prices of a variety of 
derivative contracts.  

More generally, one can envisage a market admitting numerous decentralized currencies. Now, in 
a friction-free market with $n$ cryptocurrencies, if we write $S_t^{ij}$ ($i,j = 1, 2, 3, \dots, n)$ for 
the price at time $t$ of one unit of currency $i$  quoted in units of currency $j$, then we have 
\begin{eqnarray} 
S_t^{ij}=\pi_t^i \, / \pi_t^j \, , 
\end{eqnarray}
where $\{\pi_t^i\}$ denotes the pricing kernel for currency $i$  
\cite{FH,Rogers, Lipton}. In the present modelling framework, one can imagine a situation in which 
the pricing kernels each take the form (\ref{pi}), with initial values 
$\pi_0^i=(a_i^2+b_i^2+c_i^2)^{-1/2}$ and respective $\{\sigma_t^i\}$ functions. Some of the 
Brownian motions are shared throughout the crypto economy, representing systematic risk, while 
others may apply, perhaps, only to one or two cryptocurrencies, representing idiosyncratic risk. A call 
option on the crypto exchange rate with maturity $T$ and strike rate $K$ will have the payout 
$H_T=(\pi_T^i/\pi_T^j-K)^+$, the price of which can easily be computed numerically 
on account of the Gaussian nature of the setup. 

In a similar vein, one can examine the problem of pricing a call option on the exchange rate between 
cryptocurrency $i$ and a sovereign currency, say, USD. Then the dollar price of an option to purchase
one unit of cryptocurrency $i$  at time $T$ at the strike price $K$ is
\begin{eqnarray} 
C_0^{i\$}= \frac{1}{\pi_0^\$} \, {\mathbb E}\left[(\pi_T^i-K\pi_T^\$)^+\right].
\end{eqnarray}
As an illustration, suppose that we have a geometric Brownian motion model 
\begin{eqnarray} 
\pi_t^\$ = \pi_0^\$ \, \re^{-rt -\lambda B_t - 
\frac{1}{2}\lambda^2 t}
\label{GBM}
\end{eqnarray}
for the dollar pricing kernel, where the
short rate $r$ is constant. For the cryptocurrency (say, bitcoin), we consider a pricing kernel of 
the form (\ref{pi}) with initial value $\pi_0^{\bitcoinB}=(a^2+b^2+c^2)^{-1/2}$, and such that $\{B_t\}$ and $\{W_t\}$ are independent. The option price is then determined by the expectation of the random variable 
$(\pi_T^{\bitcoinB}-K\pi_T^\$)^+$, which is nonzero only if 
\begin{eqnarray}
B_T  > - \frac{1}{\lambda} \, \left[ \log \left( \pi_T^{\bitcoinB}/K\pi_0^\$ \right)+ rT + \frac{1}{2}\lambda^2 T \right]. 
\end{eqnarray}
A straightforward calculation then shows that the option price is given by
\begin{eqnarray}
C_0^{{\bitcoinB}\$} = \frac{1}{\pi_0^\$} \, {\mathbb E}\left[ \pi_T^{\bitcoinB} N(g_+) - K\re^{-rT}\pi_0^\$ N(g_-) \right] , 
\end{eqnarray}
where 
\begin{eqnarray}
g_{\pm} =  \frac{  \log \left( \pi_T^{\bitcoinB}/K\pi_0^\$ \right) + rT \pm \frac{1}{2}\lambda^2 T}
{\lambda\sqrt{T}}  
\end{eqnarray}
and
\begin{eqnarray}
N(x) = \frac{1}{\sqrt{2\pi}} \, \int_{- \infty}^x {\re}^{-\half z^2} \, \rd z \, .
\end{eqnarray}
Hence, the cryptocurrency exchange-rate option prices can easily be computed numerically. Although for simplicity here we have taken the dollar term structure to be exponential with a constant short rate, it is straightforward to extend the model to allow for calibration to the initial dollar term structure, while preserving the overall tractability of the results. In this respect, our analysis of options on crypto exchange rates can be contrasted with the pioneering work of Madan, Reyners \& Schoutens \cite{MRS}, where the dollar and bitcoin interest rates are taken to be constant (in fact, zero), an assumption that is probably justifiable for the relatively short dated options currently available on the BTC-USD rate, though not very satisfactory, needless to say, from a broader perspective. With the development of interest rate models applicable to bitcoin and other cryptocurrencies, financial institutions will be in a position to trade in cryptocurrency interest rate products. It could be that distributed ledger technologies will eventually find a way of rewarding the holders of positions in virtual currencies with interest on a continuous basis. In the meantime, there should be a role for models of no interest.\\

\noindent \textbf{Acknowledgements}
The authors wish to express their gratitude to J.~R.~Boland, M.~Kecman and A.~Rafailidis for helpful comments. \\


\vspace{1.0cm}

\end{document}